\documentstyle[12pt,epsf]{article}
\begin{document}
\begin{center}
{\Large \bf Analytic model of a Regge trajectory in the space-like and time-like
regions\footnote{Presented at the Bogolyubov Conference, Moskow-Dubna-Kiev, 
27 Sept.-6 Oct., 1999.}}

\bigskip
{\large R.~Fiore$^a$, L.L.~Jenkovszky$^b$, 
V.~Magas$^{b,c}$, F.~Paccanoni$^d$, A.~Papa$^a$ }
\date{}
\smallskip

{\it

$^{a}$ Dipartimento di Fisica, Universit\`a della Calabria, \\
Istituto Nazionale di Fisica Nucleare, Gruppo collegato di Cosenza \\
I-87036 Arcavacata di Rende, Cosenza, Italy

\smallskip

$^{b}$ Bogolyubov Institute for Theoretical Physics, \\
Academy of Sciences of Ukraine \\
252143 Kiev, Ukraine

\smallskip

$^{c}$ Department of Physics, Bergen University, \\
Allegaten 55, N-5007, Norway

\smallskip

$^{d}$ Dipartimento di Fisica, Universit\`a di Padova, \\
Istituto Nazionale di Fisica Nucleare, Sezione di Padova \\
via F. Marzolo 8, I-35131 Padova, Italy}

\end{center}

\begin{abstract}
A model for a Regge trajectory compatible with the threshold behavior 
required by unitarity and asymptotics in agreement with Mandelstam
analyticity is analyzed and 
confronted with the experimental data on the spectrum of the $\rho$ 
trajectory as well as those on the $\pi^- p\rightarrow \pi^0 n$ charge-exchange 
reaction. The fitted 
trajectory deviates considerably from a linear one both in the space-like and 
time-like regions, matching nicely between the two.

\end{abstract}

Regge trajectories may be considered as building blocks in the framework of 
the analytic $S$-matrix theory. We dedicate this contribution to the late 
N.N.~Bogolyubov, whose contribution in this field is enormous, on the occasion 
of his 90-th anniversary. The model to be presented is an example of the  
realization of the ideas of the analytic $S$-matrix theory.  

There is a renewed interest in the studies of the dynamics of the Regge 
trajectories~\cite{Burak,FPS,Dey}. There are various reasons for this 
phenomenon. 

The hadronic string model (see e.g.~\cite{BN}) was successful as 
a mechanical analogy, generating a spectrum similar to that of a 
linear trajectory, but it fails to incorporate the interaction between the 
strings. Although intuitively it seems clear that hadron production corresponds
to breakdown of  the strings, the theory of interacting strings faces many 
problems. Paradoxically, the final goal of the hadronic string theory and, in 
a sense, of the modern strong interaction theory, is the reconstruction of the 
dual (e.g. Veneziano) amplitude from the interacting strings, 
originated by the former.

Non-linear trajectories were derived also from potential models. The 
saturation of the spectrum of resonances was shown~\cite{PST} to be 
connected to a screening quark-antiquark potential. 

A relatively new development is that connected with various quantum 
deformations, although the relation between $q$ 
deformations and non-linear (logarithmic) trajectories was first 
derived by Baker and Coon~\cite{BC}. $q$-deformations of the dual 
amplitudes (or harmonic oscillators) resulted~\cite{CK,JKM} in deviations 
from linear trajectories, although the results are rather ambiguous. By a 
different, so-called $k$-deformation, the authors~\cite{Dey} arrived at rather
exotic hyperbolic trajectories.

All these developments were preceded by earlier studies of general 
properties of the trajectories~\cite{GP}, 
that culminated in classical papers of 
the early 70-ies by E.~Predazzi and co-workers~\cite{Predazzi}, followed by 
the paper of late A.A.~Trushevsky~\cite{T}, who were able to show, on quite 
general grounds, that the asymptotic rise of the Regge trajectories cannot 
exceed $|t|^{1/2}$. This result, later confirmed in the framework of dual 
amplitudes with Mandelstam analyticity~\cite{BCJK}, is of fundamental 
importance. Moreover, wide-angle scaling behavior of the dual amplitudes
imposes an even stronger, logarithmic asymptotic upper bound on the 
trajectories. The combination of a rapid, nearly linear rise at small $|t|$ 
with the logarithmic asymptotics may be comprised in the following form of 
the trajectory:
\begin{equation}
\alpha(t)=\alpha(0)-\gamma\ln(1-\beta t),
\end{equation}
where $\gamma$ and $\beta$ are constants. 
 
The threshold behavior of the trajectories is constrained by unitarity:
\begin{equation}
{\cal I}m\,\alpha_n(t)\sim (t-t_n)^{{\cal R}e\,\alpha(t_n)+1/2},
\end{equation} 
where $t_n$ is the mass of the $n-$th threshold. The combination of this 
threshold behavior with the square-root and/or logarithmic behavior is far 
from trivial, unless one assumes a simplified square root threshold 
behavior that, combined with the logarithmic asymptotics, results in the 
following form~\cite{FJMP}
\begin{equation}
\alpha(t)=\alpha_0-\gamma\ln(1+\beta\sqrt{t-t_n}).
\label{single_traj}
\end{equation}

The next question is how do various thresholds enter the trajectory. 
In a long series of papers N.A.~Kobylinsky with his co-workers~\cite{Kobyl} 
advocated the additivity idea
\begin{equation}
\alpha(t)=\alpha(0)+\sum_n\alpha_n(t),
\label{add_traj}
\end{equation}        
with $\alpha_n(t)$ having only one threshold branch point on the physical 
sheet. The choice of the threshold masses is another controversial problem. 
Kobylinsky {\sl et al.}~\cite{Kobyl} assumed that the thresholds are made 
only of the lowest-lying particles (and their antiparticles), appearing in the 
$SU(3)$ octet and decuplet - $\pi$ and $K$ mesons and baryons ($N$, 
eventually $\Sigma$ and/or $\Xi$). We prefer to include the physical 
$4m_{\pi}$ threshold, an intermediate one at $1$ GeV, as well as a heavy one
accounting for the observed (nearly linear) spectrum of resonances on the 
$\rho$ trajectory. The masses of the latter will be fitted to the data.    

Fig.~1 shows the Chew-Frautschi plot with the trajectory (3), (4) and 
four thresholds~\cite{Kobyl} included. This trajectory matches well with 
the scattering data~\cite{Apell77,Barnes76,Binon81}, as shown in Fig.~2, 
where fits to the scattering data based on the model~\cite{AC} are presented.
 
The construction of a trajectory with a correct threshold behaviour 
and Mandelstam analyticity, or its reconstruction from a dispersion 
relation is a formidable challenge for the theory. 
This problem can be approached by starting from the following simple 
analytical model where the imaginary part of the trajectory is chosen 
as a sum of terms like
\begin{equation}
{\cal I}m\,\alpha_n(t)=\gamma_n \left( \frac{t-t_n}{t} 
\right)^{{\cal R}e\,\alpha(t_n)+1/2} \theta(t-t_n).
\end{equation}
A rough estimate of $Re\,\alpha(t_n)$  can be obtained from a linear 
trajectory adapted to the experimental data.
We have checked this approximation {\it a posteriori} and found that it works. 
It could be improved by iterating the zeroth order approximation.
From the dispersion relation for the trajectory, the real part can be 
easily calculated~\cite{BAT}
\begin{displaymath}
{\cal R}e\,\alpha(t)=\alpha(0)+\frac{t}{\sqrt{\pi}} \sum_n 
\gamma_n\frac{\Gamma(\lambda_n 
+3/2)}{\sqrt{t_n}\Gamma(\lambda_n+2)}\,{}_2F_1\left(1,1/2;\lambda_n+2;
\frac{t}{t_n}\right)\theta(t_n-t)+
\end{displaymath}
\begin{equation}
+\frac{2}{\sqrt{\pi}}\sum_n\gamma_n \frac{\Gamma(\lambda_n+3/2)}{\Gamma
(\lambda_n+1)}\sqrt{t_n}\;{}_2F_1\left(-\lambda_n,1;3/2;
\frac{t_n}{t}\right)\theta(t-t_n),
\end{equation}
where $\lambda_n={\cal R}e\,\alpha(t_n)$. Work in this direction is in progress.

\bigskip

{\bf Acknowledgment}

\noindent One of us (L.L.J.) is grateful to the Dipartimento di Fisica 
dell'Uni\-ver\-si\-t\`a 
della Calabria and to the Istituto Nazionale di Fisica Nucleare - 
Sezione di Padova e Gruppo Collegato di Cosenza for their warm hospitality 
and financial support.

\begin{figure}[htb]

\begin{center}
\begin{minipage}{75mm}
\epsfxsize=70mm
\epsfbox{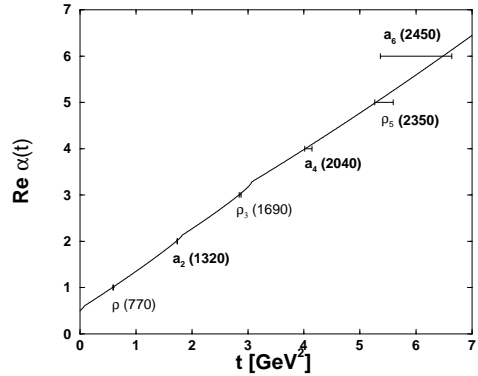}
\end{minipage}
\end{center}
\caption{Chew-Frautschi plot for the six low-lying $I=1$ parity even mesons 
($\rho$-trajectory). The masses of the resonances were taken from~\cite{Caso}.}

\end{figure}

\begin{figure}[htb]

\begin{center}
\begin{minipage}{75mm}
\epsfxsize=70mm
\epsfbox{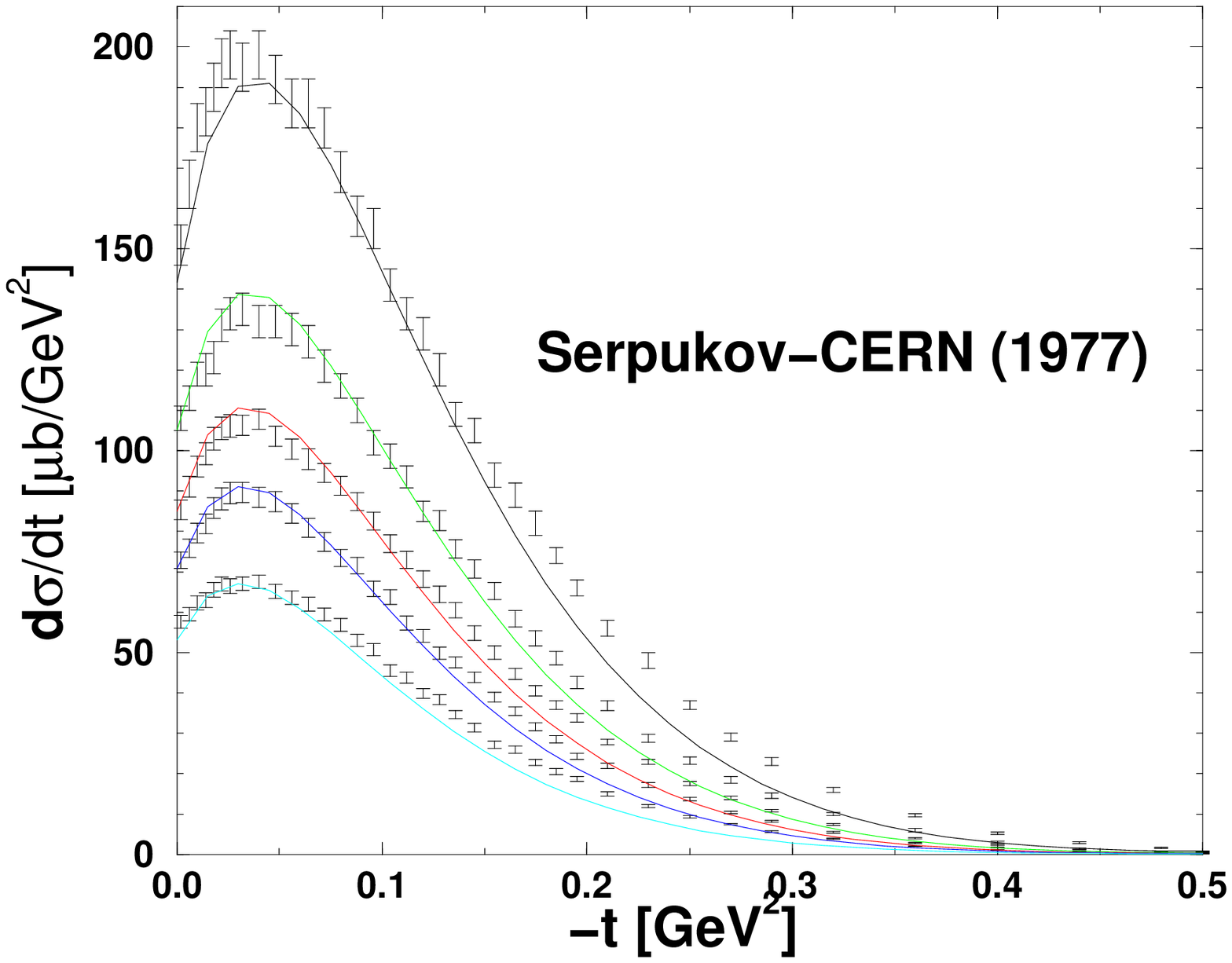}
\end{minipage}
\begin{minipage}{75mm}
\epsfxsize=70mm
\epsfbox{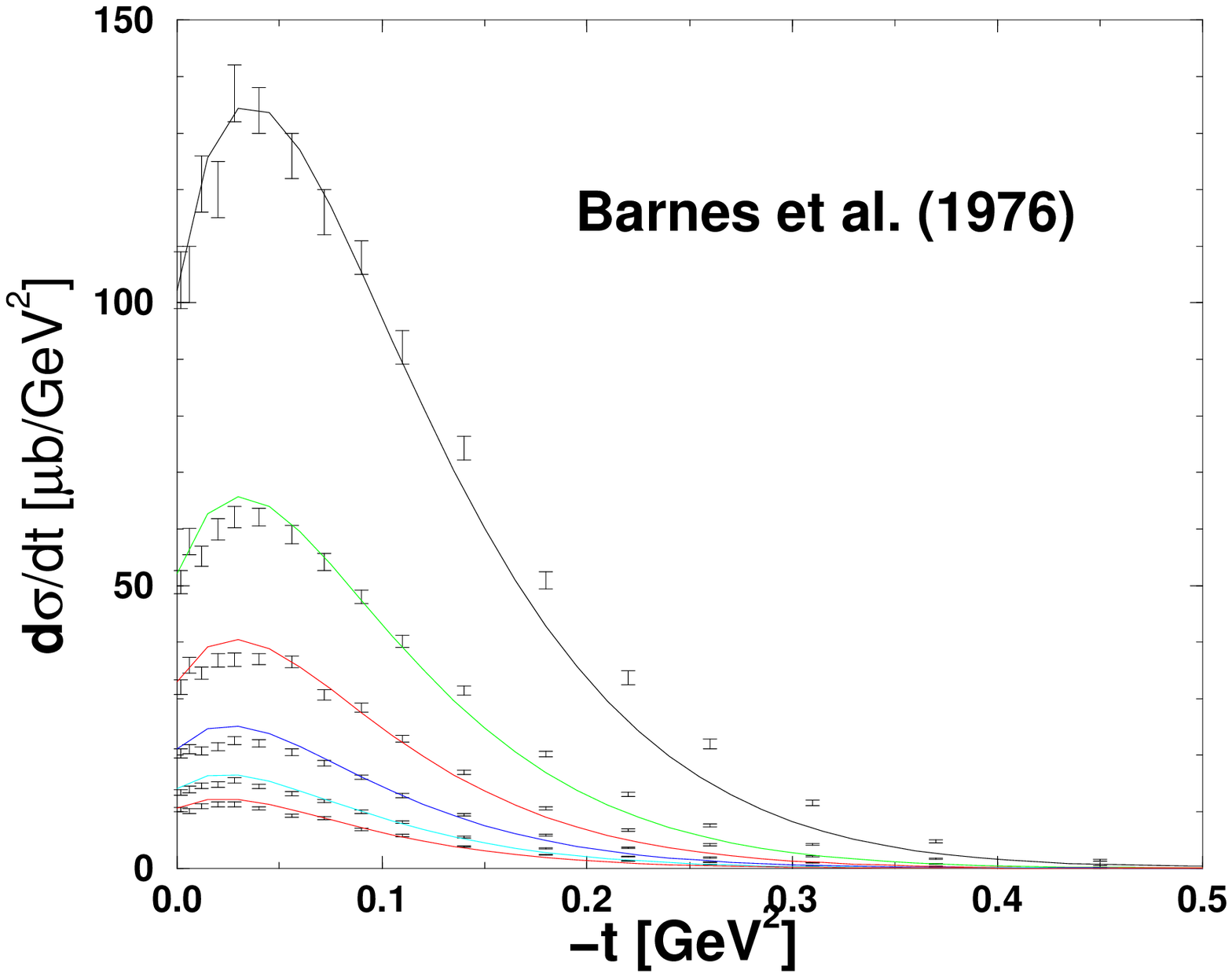}
\end{minipage}
\end{center}

\phantom{.}

\begin{center}
\begin{minipage}{75mm}
\epsfxsize=70mm
\epsfbox{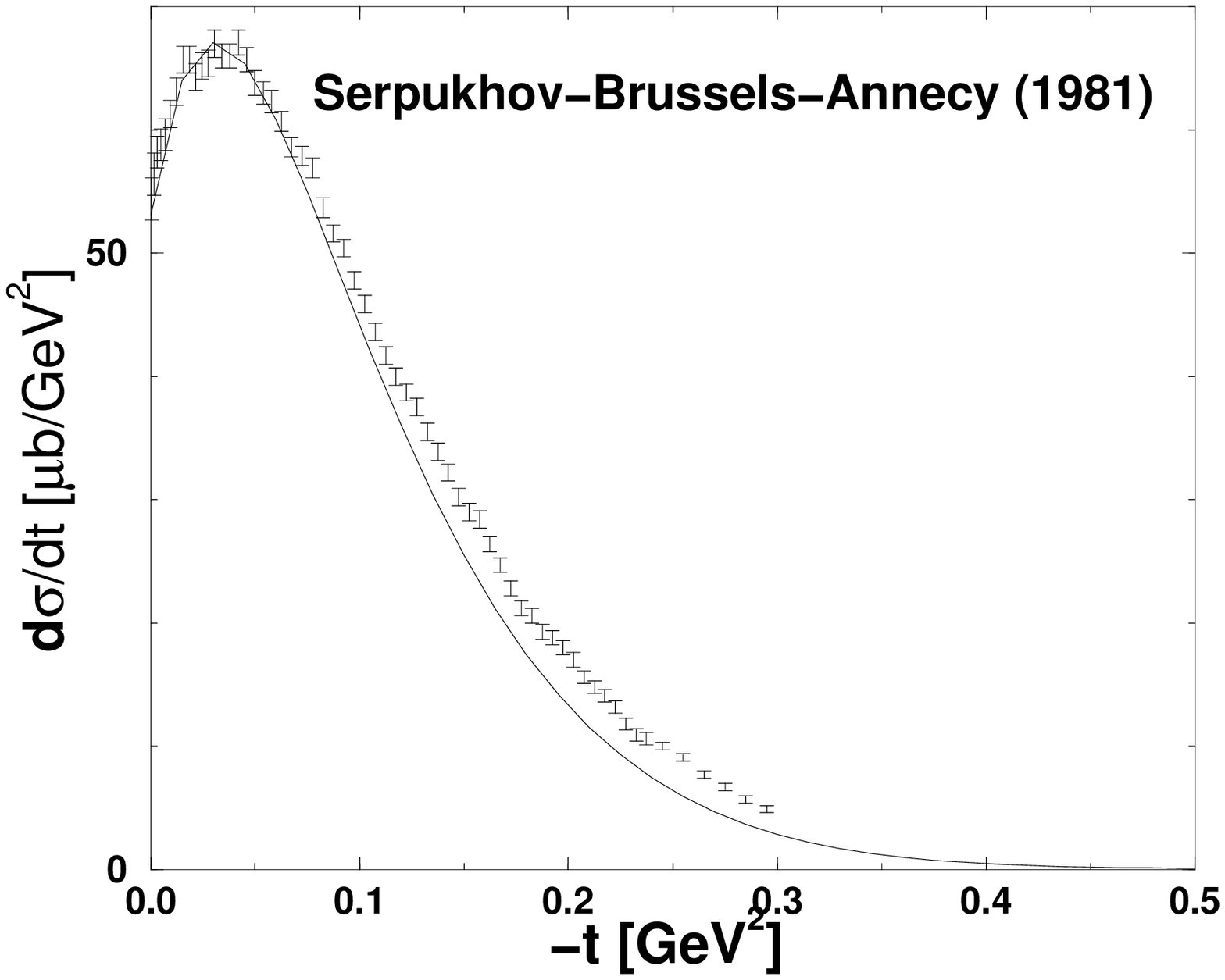}
\end{minipage}
\end{center}

\caption{Differential cross section $d\sigma/dt$ [$\mu$b/GeV$^2$]
versus $-t$ [GeV$^2$] for the process $\pi^- p \rightarrow \pi^0 n$. The 
solid curves represent the result of the fit with the model by Arbab and 
Chiu~\cite{AC} using the trajectory defined in Eqs.~(\ref{single_traj}) 
and (\ref{add_traj}). Data are taken from Ref.~\cite{Apell77} (top-left),
Ref.~\cite{Barnes76} (top-right) and Ref.~\cite{Binon81} (bottom).}

\end{figure}

\end{document}